# Structural, Electrical, and Magnetic Properties of $La_2Ni_{1.5}Mn_{0.5}O_6$ double perovskites

MOHD NASIR[1], A. K. BERA[3], S. M. YUSUF[3], N. PATRA[4], D. BHATTACHARYA[4], SUNIL KUMAR[2], S. N. JHA[4], SHUN-WEI LIU[5], SAJAL BIRING[5] and SOMADITYA SEN[1,*]

[1] Department of Physics, Indian Institute of Technology Indore, 453552, India
[2] Metallurgical Engineering and Material Science, Indian Institute of Technology Indore, 453552, India
[3] Solid State Physics Division, Bhabha Atomic Research Centre, Mumbai, 400085, India
[4] Atomic & Molecular Physics Division, Bhabha Atomic Research Centre, Mumbai, 400085, India
[5] Ming Chi University, New Taipei City 24301, Taiwan

*Corresponding author: sens@iiti.ac.in

Among multifunctional double perovskite oxides, $La_2NiMnO_6$ has recently drawn significant attention due to its importance both in terms of understanding of fundamental physics and potential for device applications. The relative alteration in Ni:Mn ratio strongly influences the structural, magnetic and electrical properties of $La_2NiMnO_6$. In the present study, $La_2Ni_{1.5}Mn_{0.5}O_6$ sample with Ni:Mn = 3:1 ratio has been prepared using sol-gel method and modifications of the above physical properties from that of a stoichiometric sample of $La_2NiMnO_6$ are discussed. Rietveld analysis of X-ray diffraction data shows that $La_2Ni_{1.5}Mn_{0.5}O_6$ samples belong to a major monoclinic structural phase ($P2_1/n$), with partially disordered arrangement of Ni and Mn ions. Magnetic characterization and X-ray absorption near edge structure analysis reveal ferromagnetic ordering around 260 K between $Ni^{2+}$ and $Mn^{4+}$ ions while spin glass like behavior is revealed at ~45 K. Semiconducting nature of the sample was confirmed from temperature dependent resistivity measurements as well as from the I-V characteristics measured at room temperature.

## I. INTRODUCTION

Double perovskites $AA'BB'O_6$ (where A and A' are rare-earth ions and $B$ and $B'$ are different transition-metal ions) are interesting transition-metal oxides due to their diverse physical properties [1-2]. Compared to simpler perovskites $ABO_3$, the presence of chemical order of $B$ and $B'$ cations influences both electronic and the magnetic structures considerably [3]. Double perovskite oxides exhibit naturally ordered structures consisting of cubic, corner connected network of alternating $BO_6$ and $B'O_6$ octahedra in all three directions [4]. Large differences in oxidation state and ionic radii between $B$ and $B'$ cations favors ordering of various types of $BO_6$ octahedrons [3].

$La_2NiMnO_6$ is a double-perovskite with rich electronic and magnetic properties. Hence it has potentials for industrial applications [2, 5-9]. It is a ferromagnetic semiconductor with high magnetic ordering temperature (Tc~280K) [10]. It shows giant magneto-dielectricity around room temperature due to B-site ordering of $Ni^{2+}$ and $Mn^{4+}$ cations [11]. Resistivity and dielectric properties of $La_2NiMnO_6$ can be varied by the magnetic field or vice versa [5,11], which make this material an exciting candidate for commercial device applications. This material has been highly utilized in solid-state thermoelectric coolers [5]. Kanamori–Goodenough ferromagnetic super-exchange interaction is a suggested mechanism in $La_2NiMnO_6$ [12-13]. X-ray absorption and X-ray/neutron diffraction studies on $La_2NiMnO_6$ provided evidences for B-site ($Ni^{2+}$ and $Mn^{4+}$ ions) long range ordering in agreement with monoclinic ($P2_1/n$) or rhombohedral ($R\bar{3}m$) symmetry [5, 7, 11, 14-15]. The two phases coincide over a broad temperature range [7]. Substitution at B-site cations can alter structural and other physical properties [16-19]. $La_2NiMnO_6$ with Ni:Mn = 1:1 ratio [8, 10, 20-24] is well studied, however, $La_2Ni_{1.5}Mn_{0.5}O_6$ with a Ni:Mn = 3:1 ratio has not been synthesized yet as per our knowledge. $La_2N_{1.5}Mn_{0.5}O_6$ compared with $La_2NiMnO_6$, has higher conductivity emphasizes the role of excess Ni content, and promotes its potential applicability. Magnetic ordering may be influenced by the excess Ni content due to changes in Ni-O-Mn interaction, regenerating the magnetic behavior of the new composition. Herein, we discuss structural, electrical and magnetic properties correlation in $La_2Ni_{1.5}Mn_{0.5}O_6$ (LNMO) double perovskite.

## II. EXPERIMENTAL

Polycrystalline $La_2Ni_{1.5}Mn_{0.5}O_6$ samples were synthesized by sol-gel assisted combustion method. Solutions of $La(NO_3)_3.6H_2O$ (Alfa Aesar, purity 99.99%), $Ni(NO_3)_2.6H_2O$ (Alfa Aesar, 99.99085% metal basis), and $Mn(NO_3)_2$ (50% w/w aq. sol., Alfa Aesar) were prepared and used as the starting materials. The solutions were mixed together and stirred vigorously for uniform mixing. A solution of citric acid and ethylene glycol were added later. The solution was heated at 70°C for 5 h with continued stirring. Once the gel was formed, heating at 100°C yielded a black fluffy powder. This powder was further ground and calcined in air for 6 hours at temperatures ~600, 900 and finally 1150°C.

The phase identification of these materials was investigated by powder X-ray diffractometer (Bruker D2 Phaser X-ray diffractometer) with Cu $K_\alpha$ as the source ($\lambda = 1.54$ Å). The spectra were collected in the $2\theta$ range, 20° - 80°. The microstructure was studied by field emission scanning electron microscopy (Carl Zeiss FESEM Supra-55). X-ray absorption near edge structure (XANES) has been carried on at Ni and Mn K-edges at the Energy Scanning EXAFS beamline BL-09 (Indus-2, RRCAT, India) [25],[26]. EXAFS beamline operates in the energy range of 4-25 KeV with a resolution of 1 eV at 10 KeV. The beamline uses a Si (111) double-crystal monochromator to select the energy of the X-ray beam from synchrotron light produced by electron storage ring (2.5 GeV, $I_{max}$~300 mA). XANES spectra of LNMO were recorded at room temperature at Ni (8333 eV) and Mn (6539 eV) $K$-edges in fluorescence mode. Standard metal foils of Ni and Mn were used for each energy calibration. A vibrating sample magnetometer (VSM) was used to measure zero field-cooled (ZFC) and field-cooled (FC) temperature dependent dc magnetization in the temperature range 5-300 K at 250 Oe magnetic field. Temperature dependent ac susceptibility at various frequencies was measured for T~5-300K using a commercial ac-susceptibility probe. Cylindrical pellets of diameter ~13 mm and thickness ~1.2 mm were prepared pressing a mixture of pure LNMO powders and 5% polyvinyl alcohol. Dense LNMO ceramics were obtained by sintering at 1500°C for 6 h. For electrical measurements, pellets were polished on both sides and cut into rectangular parallelepiped geometry. Resistivity measurements were carried out using the standard two-probe method. The conducting silver paste was employed to make electrical contacts.

## III. RESULTS AND DISCUSSION

X-ray diffraction spectra of $La_2Ni_{1.5}Mn_{0.5}O_6$ confirms single phase [Figure 1]. The XRD pattern can be indexed to monoclinic perovskites structure with no extra secondary phase [16, 27].

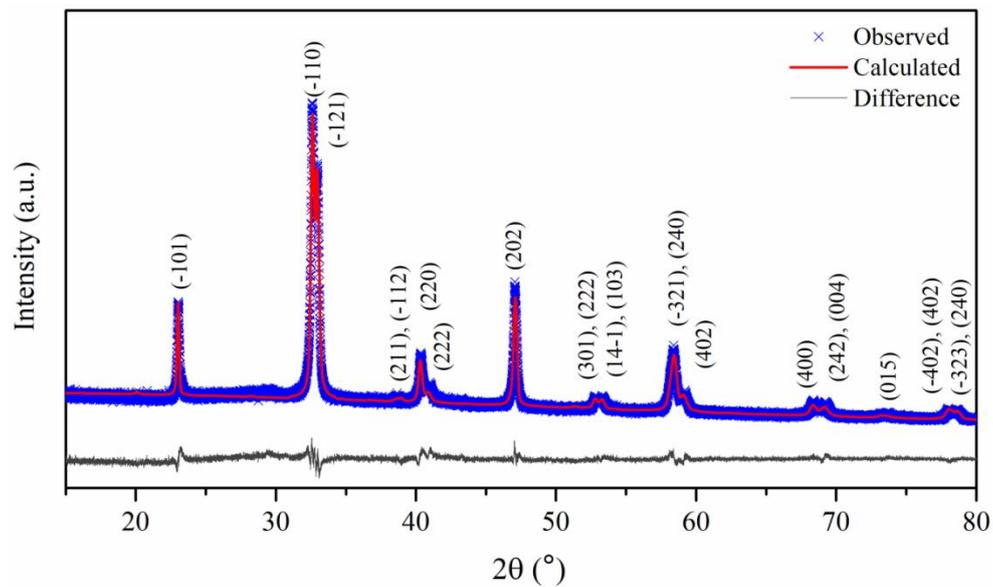

*Fig. 1-Room temperature X-ray diffraction spectrum with Rietveld refinement of $La_2Ni_{1.5}Mn_{0.5}O_6$, Cross symbols (blue color): experimental data points; Solid line (red color): Simulated fitted curve.*

Rietveld refinement was carried out using *GSAS* software with P2₁/n space group as starting structural model. Low values of Rietveld refinement parameters such as $R_{wp}$ and $R_p$ ($R_{wp}$ = 7.66%, $R_p$ = 6.16%) stress on P2₁/n as the proper space group with partially disordered arrangement of $Ni^{2+,3+}/Mn^{4+}$ ions and ensure a satisfactory fit [Figure 1]. Refined lattice parameters, $a$ = 5.423137(96) Å, $b$ = 5.487351(10) Å, $c$ = 7.731753(98) Å, and cell volume (V) = 230.066(12) Å³ and are smaller than that of $La_2NiMnO_6$ reported previously [16, 28]. The temperature dependent dc magnetization (M-T) curves [Figure 2a] of $La_2Ni_{1.5}Mn_{0.5}O_6$ reveal a ferromagnetic-to-paramagnetic transition at ~ 260 K with an anomaly at ~ 45 K.

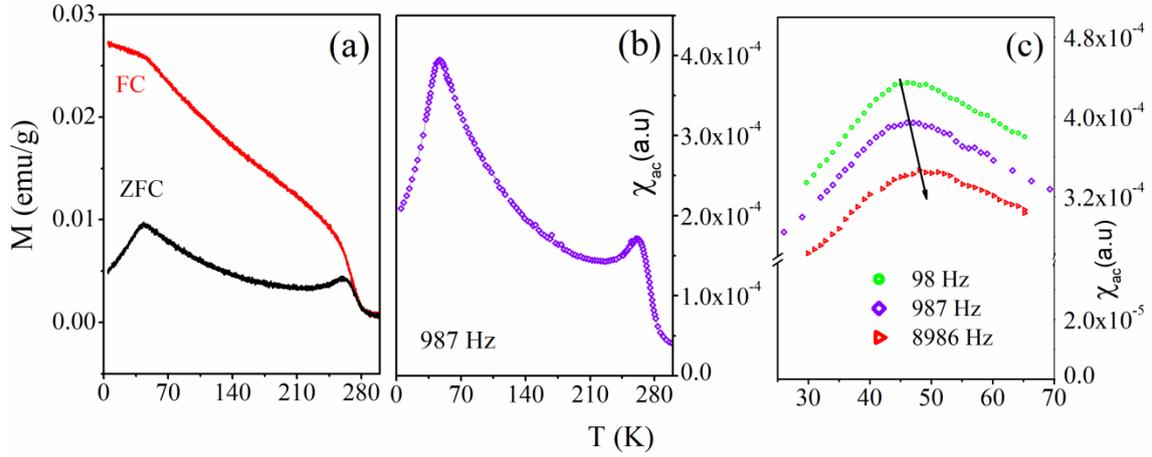

Fig. 2-*Temperature dependent (a) FC and ZFC dc magnetization under 250 Oe applied magnetic field (H), (b) Real part of ac magnetic susceptibility ($\chi_{ac}$) curve at 987 Hz, showing a magnetic transitions at 260 K and an anomaly at 46 K and (c) Real ac susceptibility ($\chi_{ac}$) at various fixed frequencies (H = 5 Oe) hinting towards the presence of a glassy state around 45 K.*

These two magnetic transitions are also present in the ac susceptibility ($\chi_{ac}$) curve [Figure 2b]. The presence of a peak in the real part of $\chi_{ac}$ at ~ 260 K indicates the onset of magnetic ordering in LNMO corresponding to $Ni^{2+}$-$O^{2-}$-$Mn^4$ super-exchange interaction [5, 11, 29]. Another peak appearing around 46 K has been interpreted in literature as either a signature of spin glass [11] or the presence of another ferromagnetic phase consisting of $Ni^{3+}$-$Mn^{3+}$ ions for a partially disordered LNMO [7]. For highly ordered La$_2$NiMnO$_6$ a single ferromagnetic transition was observed at ~ 270 K without the presence of any other magnetic transition at lower temperatures [30]. Magnetic properties give a qualitative information regarding B-site ordering, i.e. Ni and Mn ions. The low temperature magnetic transitions can be inspected using neutron diffraction, XMCD, etc. [11, 31].

XANES was employed to investigate the oxidation and spin states of the compositional elements as these properties influence magnetism [32-35]. Figure 3 shows normalized XANES spectra at Mn and Ni $K$-edges along with standard Mn foil (Mn$^0$), MnO (Mn$^{2+}$), Mn$_2$O$_3$ (Mn$^{3+}$), MnO$_2$ (Mn$^{4+}$) and Ni foil (Ni$^0$), NiO (Ni$^{2+}$) spectra respectively. By comparing the absorption energies of Ni/Mn in LNMO with standard samples, we can directly estimate the charge states of Ni and Mn ions in LNMO. It is observed that threshold energies of the Mn K-edge in LNMO are quite different from MnO and Mn$_2$O$_3$ but close to that of MnO$_2$. The Mn $K$-edge spectrum matches well with the standard MnO$_2$ spectrum [Figure 3a] revealing a majority of Mn$^{4+}$ ions in LNMO in agreement with our XRD results. The pre-edge features at Mn K-edge consist of two comparable features having partial having partial 3$d$ character (labeled as A$_1$ and A$_2$). These features are mostly realized for transition metals.

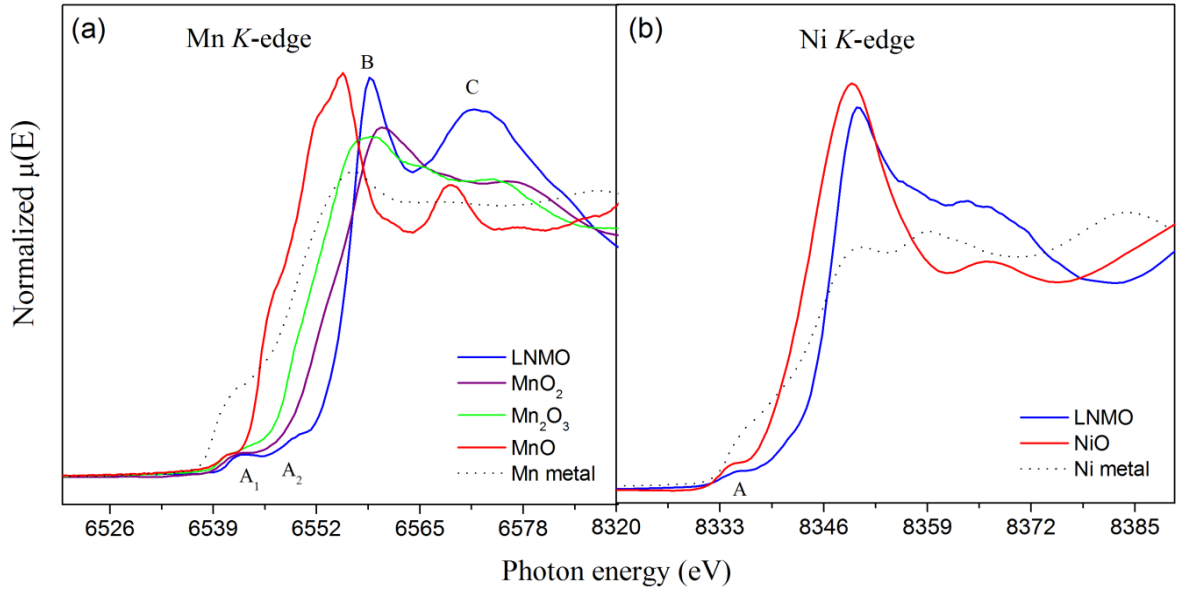

*Fig. 3-XANES spectra at (a) Mn and (b) Ni K-edges compared with standard MnO₂ and NiO spectra respectively, suggesting Mn⁴⁺ and Ni²⁺,³⁺ ions in La₂Ni₁.₅Mn₀.₅O₆. A is the pre-edge, B the main edge feature (white line) and C the shoulder attributed to multiple scattering effects.*

The pre-edge features are usually attributed to the mixing of $1s{\rightarrow}3d$ (quadrupole) and $1s{\rightarrow}3d$ (dipole) transitions. These transitions are allowed due to the hybridization between $3d$ and $4p$ states. However, $1s{\rightarrow}3d$ dipole transitions are forbidden. The splitting of the $A_1$ and $A_2$ peaks is attributed to the splitting of the majority and minority $e_g$ states [36]. Such splitting occur due to various physical effects such as strong hybridization of the $4p$ and $3d$ states of Mn atoms on adjoining sites. The feature $A_1$ arises due to the transitions into majority $e_g$ states and feature $A_2$ is attributed to transitions into minority $e_g$ states and $t_{2g}$ states [37]. The transitions, Mn $1s{\rightarrow}$Mn $4p$ states give rise to an intense peak B also called white line and multiple scattering effects are responsible for an another feature C. All these features are quite common for Mn $K$-edge on manganites. The Ni $K$-edge threshold energies of LNMO are higher than NiO, hinting at mixed valence state of +2 and +3 of Ni ions. For Ni $K$-edge, pre-edge feature A comes from $1s{\rightarrow}3d$ transitions and probes weight of Ni $3d$ states which are hybridized with Ni $4p$ states. These results directly exhibit that Ni ions are in bi-valence state and Mn ions are in tetravalent state in LNMO. This maintains charge balance as the decrease in Mn content is compensated by increase in valence of Ni, while valence state of Mn remains unchanged. Hence, XANES results combined with XRD and magnetization emphasize on the presence of a single magnetic transition in LNMO. Ac susceptibility measurements were carried out on LNMO at various frequencies, to understand the nature of low temperature magnetic transition [Figure 2c]. The transition at ~45 K shows a small shift with frequency. This is a characteristic feature of spin glass sates. In La₂Ni₁.₅Mn₀.₅O₆, half of the Mn sites are occupied by Ni. Therefore, half of the interactions are expected to be between Ni²⁺,³⁺ and Ni²⁺,³⁺. This may be related to the appearance of low temperature glassy phase. Thus LNMO consists a single ferromagnetic transition at ~260 K corresponding to Ni²⁺ and Mn⁴⁺ ordering at B site and a magnetic glassy state at lower temperature corresponding to $Ni^{2+}$-$O^{2-}$-$Ni^{2+}$/ $Ni^{3+}$-$O^{2-}$-$Ni^{3+}$ interaction. These results suggest that $La_2Ni_{1.5}Mn_{0.5}O_6$ system is partially disordered, consistent with previous reports on partially disordered $La_2NiMnO_6$ [11].

The temperature dependent resistivity of single-phase monoclinic LNMO samples was measured by four-probe technique in the range 87 K ≤ T ≤ 350 K. The sample shows a semiconducting nature with very low resistivity ~$10^{-4}$ Ω m at room temperature [Figure 4].

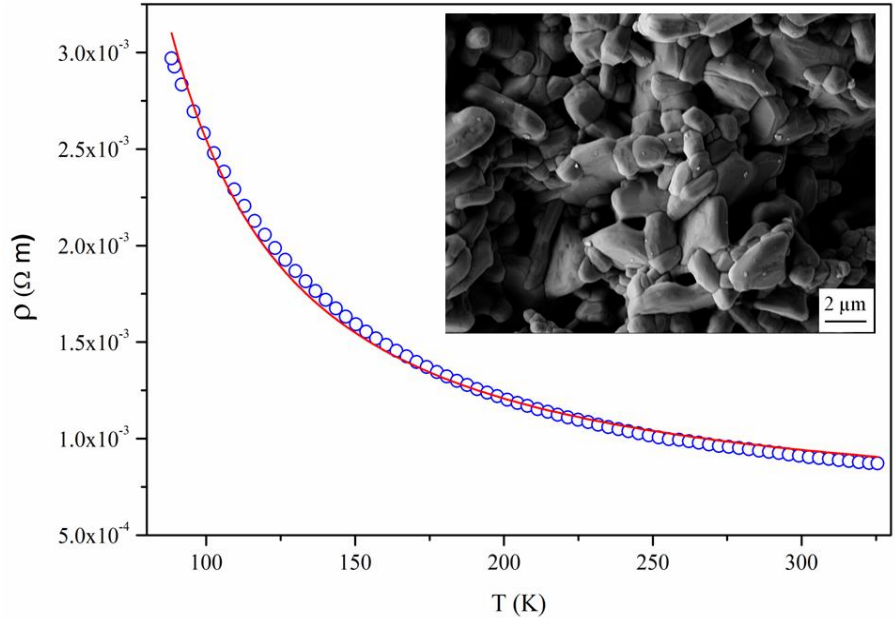

*Figure 4: Temperature dependent resistivity of $La_2Ni_{1.5}Mn_{0.5}O_6$; Red line: VRH fitting; Inset: SEM morphology for LNMO sintered at 1500ºC.*

Similar to $La_2NiMnO_6$, but with much lower values, resistivity increases with decreasing temperature without any anomaly near magnetic ordering temperature [29-30]. This decrease in resistivity is attributed to excess Ni content at B-site than Mn concentration. Thus we can tune electrical conductivity of LNMO as per our requirement to make this material an exciting candidate for practical spintronic applications. The resistivity $\rho$(T) can be fitted with variable range hopping model (VRH): $\rho(T) = \rho_0 \exp(T/T_0)^{1/4}$, where $\rho_0$ is a constant and $T_0$ is a characteristic hopping temperature which is related to electron hopping probability (P) as $T_0 \propto$ P. A satisfactory fit has been achieved as shown in Figure 4. A FESEM micrograph [the inset of Figure 4a] shows dense samples sintered at 1500ºC. The average grain size is ~2–3 μm (estimated using Image-J software). Two probe current-voltage (I-V) characteristics [Figure 5] of LNMO ceramics show a semiconducting diode nature. This may be utilized for rectifying purposes.

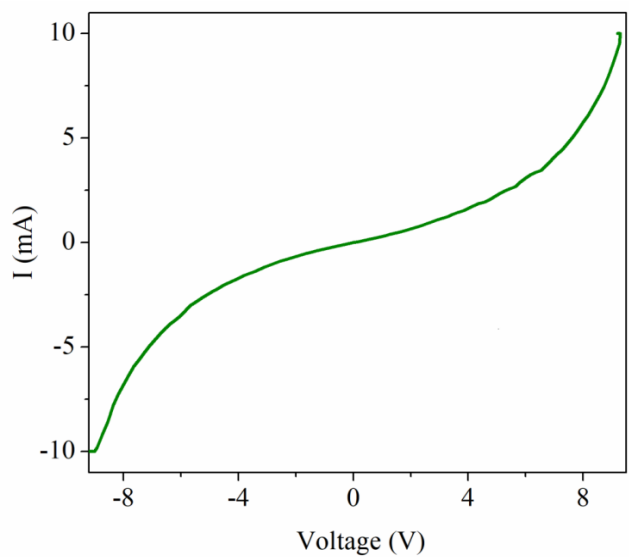

*Fig. 5-Current-voltage (I-V) characteristics of $La_2Ni_{1.5}Mn_{0.5}O_6$*

## IV. CONCLUSION

PPartially disordered $La_2Ni_{1.5}Mn_{0.5}O_6$ double perovskites were synthesized using sol-gel method. Structural, electrical and magnetic characterizations were done using a set of complementary experimental tools. Rietveld analysis reveals a phase pure monoclinic structure of $P2_1/n$ space group. XANES and magnetization results revealed ferromagnetic ordering around 260 K (attributed to the B-site ordering of $Ni^{2+}$ and $Mn^{4+}$ cations). A spin glass like feature was also noticed at lower temperature around 45 K, originating from $Ni^{2+,3+}$ and $Ni^{2+,3+}$ interaction. LNMO shows semiconducting nature with large conductivity at room temperature. Thus, the electrical and magnetic properties can be tuned with modifications in crystal structure. Tuning the B-site substitution, potential materials can be designed for practical industry applications.

## ACKNOWLEDGMENT

The authors gratefully acknowledge IIT Indore and Sophisticated Instrumentation Center, IITI, for funding the project and providing SEM analysis. The authors thank Dr. O. Okram, UGC-DAE CSR Indore for providing facility for electrical measurements. Md Nasir acknowledges financial support received from UGC, New Delhi under Maulana Azad fellowship.

# LIST OF FIGURE CAPTIONS;

Fig. 1-Room temperature X-ray diffraction patterns with Rietveld refinement of $La_2Ni_{1.5}Mn_{0.5}O_6$. Cross symbols (blue color): experimental data points; Solid line (red color): Simulated fitted curve.

Fig. 2-Temperature dependent (a) FC and ZFC dc magnetization under 250 Oe applied magnetic field (H), (b) Real part of ac magnetic susceptibility ($\chi_{ac}$) curve at 987 Hz, showing a magnetic transitions at 260 K and an anomaly at 46 K and (c) Real ac susceptibility ($\chi_{ac}$) at various fixed frequencies (H = 5 Oe) hinting towards the presence of a glassy state around 46 K.

Fig. 3-XANES spectra at (a) Mn and (b) Ni K-edges compared with standard $MnO_2$ and NiO spectra respectively, suggesting $Mn^{4+}$ and $Ni^{2+,3+}$ ions in $La_2Ni_{1.5}Mn_{0.5}O_6$. A is the pre-edge, B the main edge structure (white line) and C the shoulder associated with middle range effects.

Figure 4: Temperature dependent resistivity of $La_2Ni_{1.5}Mn_{0.5}O_6$; Red line: VRH fitting; Inset: SEM morphology for LNMO sintered at 1500 °C.

Fig. 5-Current-voltage (I-V) characteristics of $La_2Ni_{1.5}Mn_{0.5}O_6$